\documentclass[10pt, a4paper,english]{article} 
\usepackage{csquotes}
\usepackage{float}
\usepackage{graphicx}

\usepackage{amsmath, amssymb, amsthm}
\usepackage{pictex, dcpic}
\usepackage{xcolor}

\def\al{\alpha}
\def\be{\beta}
\def\de{\delta}
\def\ga{\gamma}

\def\ep{\epsilon}

\def\te{\theta}
\def\la{\lambda}
\def\ze{\zeta}

\def\si{\sigma}

\def\De{\Delta}
\def\Ga{\Gamma}

 \def\calC{{\hbox{\cal C}}}

 \def\calM{{\hbox{\cal M}}}

 \def\E{\mathbb{E}}

 \def\L{\mathbb{L}}

 \def\Q{\mathbb{Q}}
 \def\R{\mathbb{R}}

\def\Aut{{\hbox{Aut}}}

\def\Iso{{\hbox{Iso}}}

\def\GL{{\hbox{GL}}}
\def\det{{\hbox{det}}}

\def\Lor{{\hbox{Lor}}}
\def\Diff{{\hbox{Diff}}}

\def\ip{\hbox to4pt{\leaders\hrule height0.3pt\hfill}\vbox to8pt{\leaders\vrule width0.3pt\vfill}\kern 2pt}

\def\del{\partial}

\def\arr{\rightarrow}

\def\then{\Rightarrow}

\def\barJ{\bar J}

\def\calH{{{\cal H}}}

\let\Latexfrac\frac
\def\frac[#1/#2]{\hbox{$#1\over#2$}}
\def\Frac[#1/#2]{{#1\over#2}}
\def\({\left(}
\def\){\right)}
\def\[{\left[}
\def\]{\right]}
\def\^#1{{}^{#1}_{\>\cdot}}
\def\_#1{{}_{#1}^{\>\cdot}}
\def\Label=#1{{\buildrel {\hbox{\fiveSerif \ShowLabel{#1}}}\over =}}
\def\<{\kern -1pt}

\def\Uvec#1{\vbox{\mathsurround=0pt\ialign{##\crcr
     $\scriptscriptstyle\rightharpoonup$\crcr\noalign{\kern1pt\nointerlineskip}
     $\hfil\displaystyle{#1}\hfil$\crcr}}}
\def\Dvec#1{\vbox{\mathsurround=0pt\ialign{##\crcr
     $\scriptscriptstyle\rightharpoondown$\crcr\noalign{\kern-7pt\nointerlineskip}
     $\hfil\displaystyle{#1}\hfil$\crcr}}}

\def\uvecu{\vbox{\mathsurround=0pt\ialign{##\crcr
     $\scriptscriptstyle\rightharpoonup$\crcr\noalign{\kern1pt\nointerlineskip}
     $\hfil\displaystyle{u}\hfil$\crcr}}}
\def\dvecu{\vbox{\mathsurround=0pt\ialign{##\crcr
     $\scriptscriptstyle\rightharpoondown$\crcr\noalign{\kern-7pt\nointerlineskip}
     $\hfil\displaystyle{u}\hfil$\crcr}}}

\def\uvecbe{\vbox{\mathsurround=0pt\ialign{##\crcr
     \kern3pt$\scriptscriptstyle\rightharpoonup$\crcr\noalign{\kern1pt\nointerlineskip}
     $\hfil\displaystyle{\be}\hfil$\crcr}}}
\def\dvecbe{\vbox{\mathsurround=0pt\ialign{##\crcr
     \kern1pt$\scriptscriptstyle\rightharpoondown$\crcr\noalign{\kern-10pt\nointerlineskip}
     $\hfil\displaystyle{\be}\hfil$\crcr}}}

\def\uvecn{\vbox{\mathsurround=0pt\ialign{##\crcr
     $\scriptscriptstyle\rightharpoonup$\crcr\noalign{\kern1pt\nointerlineskip}
     $\hfil\displaystyle{n}\hfil$\crcr}}}
\def\dvecn{\vbox{\mathsurround=0pt\ialign{##\crcr
     $\scriptscriptstyle\rightharpoondown$\crcr\noalign{\kern-7pt\nointerlineskip}
     $\hfil\displaystyle{n}\hfil$\crcr}}}

\def\uvecm{\vbox{\mathsurround=0pt\ialign{##\crcr
     $\scriptscriptstyle\rightharpoonup$\crcr\noalign{\kern1pt\nointerlineskip}
     $\hfil\displaystyle{m}\hfil$\crcr}}}
\def\dvecm{\vbox{\mathsurround=0pt\ialign{##\crcr
     $\scriptscriptstyle\rightharpoondown$\crcr\noalign{\kern-7pt\nointerlineskip}
     $\hfil\displaystyle{m}\hfil$\crcr}}}

\def\uvecN{\vbox{\mathsurround=0pt\ialign{##\crcr
     \kern3pt$\scriptscriptstyle\rightharpoonup$\crcr\noalign{\kern1pt\nointerlineskip}
     $\hfil\displaystyle{N}\hfil$\crcr}}}
\def\dvecN{\vbox{\mathsurround=0pt\ialign{##\crcr
     \kern0pt$\scriptscriptstyle\rightharpoondown$\crcr\noalign{\kern-10pt\nointerlineskip}
     $\hfil\displaystyle{N}\hfil$\crcr}}}

\def\uvecu{\vbox{\mathsurround=0pt\ialign{##\crcr
     $\scriptscriptstyle\rightharpoonup$\crcr\noalign{\kern1pt\nointerlineskip}
     $\hfil\displaystyle{u}\hfil$\crcr}}}
\def\dvecu{\vbox{\mathsurround=0pt\ialign{##\crcr
     $\scriptscriptstyle\rightharpoondown$\crcr\noalign{\kern-7pt\nointerlineskip}
     $\hfil\displaystyle{u}\hfil$\crcr}}}

\def\uvecw{\vbox{\mathsurround=0pt\ialign{##\crcr
     $\scriptscriptstyle\rightharpoonup$\crcr\noalign{\kern1pt\nointerlineskip}
     $\hfil\displaystyle{w}\hfil$\crcr}}}
\def\dvecw{\vbox{\mathsurround=0pt\ialign{##\crcr
     $\scriptscriptstyle\rightharpoondown$\crcr\noalign{\kern-7pt\nointerlineskip}
     $\hfil\displaystyle{w}\hfil$\crcr}}}

\def\uvecv{\vbox{\mathsurround=0pt\ialign{##\crcr
     $\scriptscriptstyle\rightharpoonup$\crcr\noalign{\kern1pt\nointerlineskip}
     $\hfil\displaystyle{v}\hfil$\crcr}}}
\def\dvecv{\vbox{\mathsurround=0pt\ialign{##\crcr
     $\scriptscriptstyle\rightharpoondown$\crcr\noalign{\kern-7pt\nointerlineskip}
     $\hfil\displaystyle{v}\hfil$\crcr}}}

\def\union{\cup}

\def\barJ{\kern 3pt \bar {\kern -3pt J}}
\def\bfx{\mathbf{x}}

\def\cC{\calC}

\def\ShowLabel#1{\ref{#1}}

\def\ss{\smallskip}

\def\eq#1{\begin{equation}#1\end{equation}}
\def\eqLabel#1#2{\begin{equation}#1\label{#2}\end{equation}}

\def\Cases#1{\begin{cases}#1\end{cases}}
\def\Matrix#1{\begin{matrix}#1\end{matrix}}
\def\Align#1{\begin{aligned}#1\end{aligned}}

\def\eqs#1{\eq{\Align{#1}}}
\def\eqsLabel#1#2{\eq{\Align{#1}\label{#2}}}

\long\def\Note#1{\blockquote{\footnotesize #1}}

\date{}

\def\Figure[#1]#2{\begin{figure}[htbp] 
   \centering
   \includegraphics[#1]{#2} }

\def\EndFigure{\end{figure}}

\def\Item[#1]{\item[#1]}

\title{Lecture Notes in Loop Quantum Gravity.\\
LN2: Cauchy problems and pre-quantum states}

\author{\small S.Coriasco$^{a}$, L.Fatibene$^{a,b}$,  S.Garruto$^c$, A.Orizzonte$^{a}$\\
\\
\small$^a$ Department of Mathematics {\it``Giuseppe Peano''}, University of Torino (Italy)\\
\small$^b$ Ist. Naz. Fisica Nucleare (INFN) - Sezione Torino - Iniziativa spec. QGSKY (Italy)\\
\small$^c$ Department of Business and Management,  LUISS Guido Carli, Roma (Italy)\\
}

\begin{document}


\maketitle

\begin{abstract}
We discuss the structure of covariant equations, relating analytical properties of solutions to algebraic properties of the corresponding differential operator,
specifically of its principal symbol.
The principal symbol and its globality is discussed for a general quasi-linear PDE system, regardless the algebraic structure the configuration space can have.

We also discuss how the typical relativistic model can be under-deter\-mined and over-determined at the same time as well as how one can define out of it a well-posed Cauchy problem.
This issue leads us to pre-quantum configurations and Cauchy bubbles as the way to set up evolution problems in a compact region of spacetime, taking into account that relativistic models are defined on bare manifolds.  
The typical application we shall sketch is standard GR.
\end{abstract}

\section{Introduction}

Let us take  a break from the specific gravitational Holst model  we discussed in the previous paper (see \cite{LN1}) and discuss the generic structure of field equations in a relativistic theory.
We will be right back to Ashtekar-Barbero-Immirzi (ABI) model in the next lecture.

Here we wish  to discuss, from a slightly more  general perspective, how to extract (physically relevant) information about solutions from the algebraic structure of the field equations.
We are aware that this is not the only, nor the most popular, method to do it. 
Usually, this is done by the canonical analysis based on Hamiltonian formalism, in which one has a lot of structures which make the analysis technically easier.
However, one should consider that the Hamiltonian formalism in field theories is not a canonical framework as it is in mechanics (there are (between slightly and considerably) different approaches  to deal  with both the facts that one is using fields 
and that most dynamics in field theories are degenerate), and often (for instance, in ADM formalism) it partially breaks manifest general covariance that, eventually, needs to be, sometimes painfully, restored {\it a posteriori}.
Moreover, it has been noted (see \cite{Rovelli1}) that eventually LQG is concerned  only with the quantization of constraint equations of the ABI model, which are independent of the formalism one uses to find them.
Hamilton equations are defined to be equivalent to Euler-Lagrange equations, thus constraint equations can be found in both formalisms.

In quantum gravity, one wishes to discuss a covariant quantization based on something like Feynman's path integrals, which in the case of LQG leads to spin foams (see \cite{Rovelli2}). We believe it is interesting to discuss the structure of field equations and solution space also in a purely Lagrangian context. 
This will also give us some insight about the relation between classical and quantum models, between LQG and numerical relativity. 

\ss

Let us  consider a particular class of PDE on a bare manifold $M$, wide enough to encompass all field equations one finds in the relativistic (and gauge) theories relevant to physics.
Since in field theory we obtain field equations  from a variational principle, we start from a {\it configuration bundle} $[\pi:C\arr M]$ with fibered coordinates $(x^\mu, y^i)$.
Fields are sections of the configuration bundle, locally in coordinates they are given by $y^i(x)$. 
However, we know coordinates are chosen by (and to some extent identified with) observers and we want a description of physics and its equations which is independent of the observers.
To this aim, we need to know how the local field representations are affected when one changes the observers, namely their {\it transformation laws}. 

\Note{
As a side note, it has been sometimes claimed that fiber bundles are relevant to physics only when they are not trivial, since when they are trivial they in fact are Cartesian products.
Besides the fact that there are solutions of Maxwell equations which are global sections of non-trivial bundles, let us stress that this triviality argument is off-target.
We do not introduce bundles to describe globality; we do instead to have {\it transformation laws} in a categorical way.
Even when  trivial, the bundle structure defines transformations which are going to be symmetries and fiber bundles are a means to single out these transformations to be preserved by dynamics. 
Also on a trivial bundle as $\R^2\simeq\R\times \R$, when we single out transformations 
\eqLabel{
t'=t'(t)\qquad\qquad x'=x'(t, x)
}{BundleR2}
these are preserving the vertical $x$-direction and we are working on a bundle $[p:\R^2\arr \R]$, not on a product.
In fact, transformations (\ShowLabel{BundleR2}) do not preserve the product structure unless we restrict to $x'=x'(x)$.
If these transformations represent changes of observer, setting $x'(x)$ excludes relatively moving observers, which simply is not what we would physically do. 

Even though $T\R^4$ {\it is trivial}, in GR we do not want to confuse a component of a vector field (which is certainly not observable) with a scalar function!
An example more specific to LQG is the Barbero-Immirzi connection $A^i$ which,  
later on, will play a major role through its holonomies {\it because} holonomies are gauge-covariant quantities, which makes sense in view of transformation laws of $A$, not in view of its globality. 

Once one has transformation laws of an object, it is {\it always} possible to define a suitable bundle in which global sections are in one-to-one correspondence with the objects we are considering. 
Hence, our functional space will be the set of {\it global sections} of the configuration bundle. 
Global sections encode physical quantities because they are {\it absolute} and they do not depend in the observers, not in view of their globality, which is a geometric by-product of transformation laws.
}

We want a relativistic theory to be covariant with respect to $\Diff(M)$, hence the group $\Diff(M)$ has to act on the configuration bundle in the first place.
One can prove that one has an action $\la:\Diff(M)\arr \Aut(C)$ (which preserves composition, i.e.~$\la$ is a group homomorphism) iff $C$ is an {\it associated bundle} to the frame bundle $[p^1:L(M)\arr M]$
(or to some higher order analogous of it, denoted by $[p^s:L^s(M)\arr M]$). In that case, we say the configuration bundle is a {\it natural bundle}; see \cite{Kolar}, \cite{book1}.
Having a natural configuration bundle is pre-requisite to impose general covariance, it means one has spacetime diffeomorphisms acting on fields, which hence can be eventually required to be symmetries.
Of course, all tensor (densities) fields are sections of natural bundles, since they do transform with respect to spacetime diffeomorphisms.

\Note{
Also manifold connections $\Ga$ are natural objects, since they transform with respect to manifold diffeomorphisms, just the transformation laws depend on the Jacobian and the {\it Hessian}.
One needs a bigger group $\GL^2(m)$ and in fact manifold connections are global sections of a natural bundle associated to $L^2(M)$.
Also for gauge theories the issue is more complicated. They need a more general framework (see \cite{Kolar}, \cite{book1}) just because they have a more general group of symmetries to be encoded in the theory.
}

In the sequel, we restrict to {\it variational} field equations.
In variational calculus one starts with a Lagrangian $\L=L(j^ky) d\si$ of order $k$ (where $d\si$ is the natural local basis for $m$-forms over $M$ and we set $j^ky= (x^\mu, y^i, y^i_\mu, \dots, y^i_{\mu_1\dots \mu_k})$ to represent derivatives up to order $k$).  By pulling-back the Lagrangian along a section we get a form $(j^k\si)^\ast \L$  on $M$ which defines the action by integration on a compact domain $D$,
so that everything is manifestly independent of coordinates and observers.
Deformations $X= \de y^i\del_i$ are vertical vector fields on $C$, i.e.~sections of $[\nu:V(C)\arr C]$, which, in fact, drag configuration sections {\it deforming} them.

Field equations then are a global bundle map 
\eq{
\E : J^{2k} C \arr V^\ast(C) \otimes_C A_m(M)
}
where $V^\ast(C)$ is the dual bundle to $V(C)$ and $A_m(M)$ is the bundles of $m$-forms over $M$ (pulled back on $C$, to be precise).
That simply means that field equations are obtained by setting $\langle \E|X\rangle = e_i \de y^i=0$ for any deformation $X$, i.e.~locally
$e_i (j^{2k}y) \de y^i =0$, then $e_i (j^{2k}y)=0$.

\Note{
The setting above is purely geometric.
The aspect of some physical interest is that if fields transform as $(x'^\mu= x'^\mu(x), y'^i= Y^i(x, y))$, then field equations transform as
\eqLabel{
e'_i = \barJ \barJ^k_i e_k
}{FieldEquationTransformationLaws}
where $\barJ$ is the determinant of the inverse Jacobian of spacetime transformation $x'^\mu= x'^\mu(x)$, while $\barJ^k_i$ is the Jacobian inverse to $\Frac[\del Y^k/ \del y^i]$.

Let us repeat this once again: from a purely physical point of view, we really need to know how field equations transform, since we want to know that being a solution in one coordinate system is equivalent to be a solution in any other coordinate system.
That is just Einstein's argument which led him to restrict to tensorial equations, just slightly more general.
}

Finally, we require field equations to be {\it quasi-linear}. For first order equations, that means field equations can be written in the form
\eqLabel{
\E= e_i \bar d y^i\otimes d\si = \( \al_{ik}^\mu(x, y) \del_\mu y^k + \be_i(x, y)\) \bar d y^i\otimes d\si=0
}{FirstOrderPDE}
One can easily check that if field equations are quasi-linear in a coordinate system, they are quasi-linear in any other coordinate system, whatever field transformation $\(x'^\mu(x),Y^i(x, y)\)$ one has.
Being quasi-linear is an intrinsic property; it has to do with how partial derivatives transform, not with how fields transform. 

One can show there actually exist Lagrangians with field equations that are not quasi-linear. However, all field equations relevant to fundamental physics are (see \cite{Gelfand}, \cite{book2}).
We simply neglect {\it bad} Lagrangians with field equations that are not quasi-linear and assume we select a dynamics from {\it good} Lagrangians.

\ss
Hence, we want to discuss properties of generally covariant, quasi-linear field equations, transforming as (\ShowLabel{FieldEquationTransformationLaws}), written for global sections of a  natural configuration bundle $[\pi:C\arr M]$.
In view of the general covariance,  the so-called {\it hole argument} spoils any hope for the equations to be deterministic, i.e.~to obey some sort of Cauchy theorem-like result (see \cite{Norton}, \cite{HoleArgument}).

\Note{
Imagine one gives initial conditions on a hypersyrface $S_0\subset M$ and then looks for a solution of equations obeying initial conditions.
Since there easily are spacetime diffeomorphisms acting on $M$ but being the identity in a neighbourhood of $S_0$, and since they are symmetries because of general covariance,
one can easily write different solutions with the same initial conditions.

The same thing happens for Maxwell equations in electromagnetism (or for Yang-Mills theories), with gauge transformations acting on the potential $A$ (or $A^i$).
}

Whenever one has localized symmetries, Cauchy theorem cannot hold true. 
In particular, uniqueness is lost and equations cannot fix the evolution of all fields: 
we say field equations are {\it under-determined}, see \cite{GIMPSY}.
That is a typical feature in fundamental theories, since all of them have huge gauge groups, including plenty of localized symmetries (see \cite{HoleArgument}).
Once we know equations are under-determined, since we have as many equations as fields to be determined, it means that some of the equations do not contribute to the Cauchy problem, namely, they do not contain derivatives in the direction of evolution:
they are equations on $S_0$, i.e.~they are constraints on initial conditions.
Therefore, besides being under-determined the Cauchy problem is  (at the same time!) {\it over-determined}, in the sense we may have to drop some initial conditions which are not allowed by the constraint equations (see  \cite{GIMPSY}).

Thus, both equations and fields each split into two sets: fields split into {\it bulk fields}, which enter the {\it evolution equations}, and {\it gauge} or {\it boundary fields} that do not. 
The field equations that are not {\it bulk} (or {\it evolution}) equations are called {\it constraints} or {\it boundary equations}.
These splittings of fields and equations are always awkward and hard to get and, in any situation, one has to find a way to obtain them. 
The details do depend on the kind of fields we start from, although
eventually we get a new field theory on a spatial manifold $S$. 
That is more or less what one tries to do in numerical gravity to integrate numerically Einstein equations. 
 It is also related to what one does in LQG.
 Essentially, LQG is a (background free, non-perturbative) quantization of the  {\it boundary equations}, which, one can argue, determine the true degrees of freedom, since they determine allowed initial conditions to which a solution is  eventually associated.
 
In these lecture notes, we shall discuss how we set a Cauchy problem on a bare manifold $M$ and when Cauchy theorem holds true in the bulk (and for the bulk fields only). We shall discuss and apply it to standard, purely metric GR.
The discussion leads us quite naturally to {\it pre-quantum configurations} and pre-quantum problem, which is, more or less, a definition of a classical propagator, as a complete integral of Hamilton-Jacobi equation gives the classical propagator solving control problems in a general Hamiltonian system.

In the following lecture notes (see \cite{LN3}), we shall show that this is the main contribution (the eikonal approximation) of the quantum theory, exactly as Hamilton-Jacobi equation is the eikonal approximation
of Schr\"odinger equation.

 \section{Principal symbol and Characteristics}\label{S2}
 
 As a matter of fact, we cannot always set the value of fields at will on an arbitrary surface $S_0$, being it a Cauchy surface for initial conditions or the boundary of a compact region. 
 Depending on the system, topological conditions may have to be met by $S_0$, so that the value of fields can be set freely on it and they are not constrained by the PDE itself.
 This discussion is equivalent to proving that the problem is well-posed (on a bare manifold though) and to many physical properties (counting the physical degrees of freedom, the equations being deterministic, propagation of information, dependence regions, and so on).

 Given a first order, quasi-linear, differential operator $\E: J^1C\arr V^\ast(C)\otimes A_m(M)$ in the form (\ShowLabel{FirstOrderPDE})
 we can define the {\it principal symbol} to be the global map $\si: \pi^\ast(T^\ast M)\arr V^\ast(C)\otimes V^\ast(C)\otimes A_m(M)$ as
 \eq{
 \si(\xi) =  \xi_\mu \al^\mu_{ik}(x, y) \> \bar d y^i \otimes \bar d y^k \otimes d\si
 }
 where $\xi= \xi_\mu\> dx^\mu$ is a spacetime covector and $\pi^\ast(T^\ast M)$ is the cotangent bundle pulled-back on $C$, i.e.~a bundle with local coordinates $(x^\mu, y^i, \xi_\mu)$.
 
\Note{
 The principal symbol is a global map in view of the transformation laws of $\al^\mu_{ik}$, namely
 \eq{
\al'^\mu_{ik} = \bar J J^\mu_\nu \al^\nu_{jh} \bar J^j_i \bar J^h_k 
 }
 which are in turn a direct consequence of the transformation laws of equations (\ShowLabel{FieldEquationTransformationLaws}) and first derivatives 
 $y'^i_\mu =   J^i_k y^k_\nu \barJ_\mu^\nu +J^i_\nu  \barJ_\mu^\nu  $.

It is important to notice that, to define the principal symbol $\si$, one does not rely on assuming that the configuration bundle 
 is a vector or an affine bundle. We did not assume any algebraic structure for the sections of $[\pi:C\arr M]$. The principal symbol is associated to the quasi-linearity of the operator, which is in turn associated to the fact that 
 the jet bundle $[\pi^{k+1}_k:J^{k+1}C\arr J^kC]$
 is canonically an affine bundle for any integer $k$, whatever the bundle $C$ is.  
}
 
 Being global and intrinsic, the principal symbol is a tool to define intrinsic features of the operator $\E$. 
First of all, the operator is said to be {\it elliptic} iff, for any non-vanishing covector $\xi$, it takes values in non-degenerate bilinear forms $\Iso(V(C), V^\ast(C)) \subset V^\ast(C)\otimes V^\ast(C)$.
 
Field theories of interest for fundamental physics, especially relativistic theories, hardly ever are elliptic.
The next good thing that can happen to field equations is to be {\it hyperbolic}.



 
Let us start to give some pointwise definitions:
at a point $p=(x, y)\in C$,  we look for pairs $(V^\ast, W)$ where $V^\ast\subset T^\ast _xM$ is a subspace of dimension $(m-1)$ and $W\subset V_p(C)$ is a subspace of dimension $1$ such that
for all $\xi=\xi_\mu dx^\mu \in V^\ast$ and $w= w^i\del_i\in W$ we have
\eq{
w^i \xi_\mu \al^\mu_{ij} =0 
} 
Such a non-zero $\xi\in V^\ast$ is called a {\it characteristic covector}. 
We can also define  a 1-dimensional subspace $V= (V^\ast)^\circ = \{v: \xi(v)=0, \forall \xi\in V^\ast\}\subset T_xM$ and a non-zero $v\in V$ is called a {\it characteristic vector} at $x$
(remember we are not assuming a metric on $M$ to turn $\xi$ into a vector).
The pair $(V^\ast, W)$, as well as the pair $(V, W)$ or $(v, w)\in V\times W$, is called a {\it characteristic family} at $p\in C$.

We say an operator $\E$ is {\it strictly hyperbolic} iff it allows $n$ characteristic families $(v_i, w_i)$ with the vectors $w_i$ being independent, hence a basis of $V_p(C)$.

If we set initial conditions on a hypersurface $S_0\subset M$, we require characteristic vectors to be transverse to it.
As any hypersurface, $S_0$ defines a {\it canonical covector} for any embedding $i:S\arr M: k^a\mapsto x^\mu (k)$, namely
\eq{
\dvecu= \frac[1/(m-1)!] J^{\mu_1}_{a_1} \dots J^{\mu_{m-1}}_{a_{m-1}} \ep^{a_1\dots a_{m-1}} \ep_{\mu_0\mu_1\dots \mu_{m-1}} dx^{\mu_0}
}

\Note{
This has nothing to do with the metric, it is simply the covector which is zero iff evaluated on vectors tangent to $S_0$.
Hence, a vector $v\in T_xM$ is {\it transverse} to $S_0$ iff $\dvecu(v)\not=0$.

In adapted coordinates $(t, k^a)$, we have $\dvecu= dt$. 
Accordingly, a characteristic vector $\dot x = \al \del_0 + v^a \del_a$ is transverse to $S_0$ iff $\dvecu(\dot x)=\al\not=0$.
Hence the corresponding characteristic covectors can be found by solving $\xi_\mu \dot x^\mu=0$.
One obtains the  $(m-1)$-dimensional  subspace $V^\ast$ as 
\eq{
\xi = \la_a \xi^ a\in V^\ast
\qquad
 \xi^ a := -v^a \> dt + \al \> d k^a
}
Hence we have
\eq{
0= w^i \xi_\mu  \al^\mu_{ij}=  \la_a \( -  w^i v^a  \al^0_{ij}+  w^i  \al  \al^a_{ij}\)
\quad\then
  w^i v^a  \al^0_{ij} =   \al w^i   \al^a_{ij}
}
}

A curve $\ga:\R \arr M: s\mapsto x^\mu(x)$ is a {\it characteristics} with respect to the family $(v, w)$ iff its tangent vectors $\dot x=v\in V$ are characteristic vectors.
Then we can make a linear combination of equations in the system to obtain
\eqs{
\al w^i \al^\mu_{ij} y^j_\mu + \al w^i \be_i 
=& \al w^i \al^0_{ij} y^j_0 + \al w^i \al^a_{ij} y^j_a + \al w^i \be_i =\cr
=& w^i \al^0_{ij} \( \al  y^j_0 + v^a  y^j_a\) + \al w^i \be_i =\cr
=& w^i \al^0_{ij} \(  v^\mu  y^j_\mu  \)+ \al w^i \be_i =0
}
which in fact is an ODE along the characteristic curve associated to the characteristic vector $v$, by setting $y^j(s)= y^j(x(s))$ and one has $\dot y^j := v^\mu  y^j_\mu $.

If the system is strictly hyperbolic, we have $n$ characteristic families $(\uvecv_i , \uvecw_i )$, the vectors $\uvecw_i$ form a basis $\uvecw_i = \bar P_i^k \>\del_k$
and we get a field transformation $ \bar dY^i=P^i_k \bar dy^k$.
In adapted coordinates this corresponds to set the bilinear form $A_{ij}= \xi_a \al^a_{ij}$ in canonical form with respect to $\de_{ij}= \al^0_{ij}$.
This can be done iff there exists an $\de$-orthonormal basis $ \uvecw_i$  of eigenvectors of the eigenvalues $\la=-\xi_0$.
Hence the original symbol $\si_{ij}(\xi)$ was symmetric from the beginning, $\de_{ij}$ is non-degenerate and definite, and we can recast the original system in the form 
\eq{
\bar P^k_i \al^0_{kn} \( \al y^n_0 +   v^a  y^n_a\) 
=\bar P^k_i \al^0_{kn}  y^n_\mu \dot x^\mu 
=( \bar P^k_i \al^0_{kn}   \bar P^n_m)(  P^m_l y^l_\mu \dot x^\mu )
= -\bar P^k_i  b_k
}
which is a system of ODE $\de_{ik} \dot Y^k   = -\bar P^k_i  b_k$, 
each equation being computed along a different characteristic curve.

\Note{
Let us remark that the result is not really a decoupled system (the right hand sides can still depend on all fields $Y^k$), it is just in normal form.
Moreover, it is not even an ODE system for $n$ fields $Y^i(s)$ since, e.g., the field $Y^1(s)$ in the first equation is evaluated along the first characteristics, while if $Y^1(s)$
appears in other equations it is evaluated along a different curve, that is, it is a different function of $s$.

Still, one can set up a numerical approximation scheme to determine fields (and their derivatives up to an arbitrary order)  on the Cauchy surface $S_{t_0+dt}$
once fields are given on $S_{t_0}$, and hence one can compute derivatives along the Cauchy surface to any order.

In a sense, up to {\it ``details''} about regularities of the solution, one can see that initial conditions given on $S_0$ determine fields uniquely on nearby leaves on the evolution foliation,
provided (or until) characteristics curves do not focus or intersect more than once the leaves.
 
In other words, in order to have a well-posed global Cauchy problem, we need to ask first that characteristics intersect once and only once the Cauchy surface, namely that $M\simeq S_0\times \R$
is {\it globally hyperbolic}.
Since relativistic equations are geometrically defined and they gracefully restrict to regions, one can also define local Cauchy problems on $U\subset S_0$
and a {\it Cauchy bubble} $D\subset M$, the {\it dependency region}, in which we have $n$ characteristics curves through each point $p\in D$ and connecting $p$ back to $U$, on which the system becomes globally hyperbolic.
}

Although strictly hyperbolic systems are easy to be treated by a series of linear algebra tools, they are not the most general systems which support a well-posed Cauchy problem.
More generally we define a {\it symmetric hyperbolic} system when the principal symbol is symmetric, and $\al^0_{ij}$ is non-degenerate and definite.
We already showed that strictly hyperbolic systems are also symmetric hyperbolic. 
However, we have symmetric hyperbolic systems which are not strictly hyperbolic, while one can show that 
symmetric hyperbolic systems supports well-defined Cauchy problems (see \cite{Taylor}). 
In other words, {\it symmetric hyperbolic} encapsulates what we mean by a PDE which defines an evolution.

\Note{
To determine characteristic families, one has to solve the system $w^i \xi_\mu \al^\mu_{ij}=0$, which has non-zero solutions $w^i$ if and only if
\eqLabel{
Q(p; \xi)= \det(\xi_\mu \al^\mu_{ij}(p))=0
}{Q-Condition}
which is a homogeneous polynomial in $\xi$ of degree $n$. If this can be factorized into a product of $n$ homogeneous polynomials of degree 1 the system is strictly hyperbolic, at least when one then finds enough solutions $w^i$ for forming a basis.

However, in general,  (\ShowLabel{Q-Condition}) defines an algebraic hypersurface $C^\ast_p\subset T^\ast_x M$, which is called the {\it dual characteristic cone} (in view of the fact that, by homogeneity, if $\xi\in C^\ast_p$, then any $\la\xi\in C^\ast_p$).
In strictly hyperbolic systems, the  dual characteristic cone is the union of $n$ hyperplanes, that is why we can use linear algebra effectively, and why symmetric hyperbolic systems need extra care.
}

Let us remark the obvious fact that the dual characteristic cones $C^\ast_x$ are defined naturally in $T^\ast M$, which is canonically a symplectic manifold, and the bridge to analytic mechanics.
If we fix a solution $\si:M\arr C$ of the system, we can define a Hamiltonian
\eq{
H(x, \xi) = Q(\si(x), \xi): T^\ast M\arr \R
}
which defines a Hamiltonian system
\eqLabel{
\dot x^\mu=\Frac[\del H/\del \xi_\mu]
\qquad\qquad
\dot \xi_\mu = -\Frac[\del H/\del x^\mu]
}{HanitonianSystem}
A solution $(x(s), \xi(s))$ of such a system, starting from initial conditions $p=(x_0, \xi)\in C^\ast_{p}$, is called a {\it (lifted) characteristic curve}.
Being the Hamiltonian independent of $s$, it is a first integral and constant along solutions, since if it vanishes at initial conditions, it is zero along the whole characteristic curve.
In other words, characteristic curves are contained into the hypersurface $H=0$.
The tangent vectors $(\dot x, \dot \xi)\in T(T^\ast M)$ to (lifted) characteristic curves are called {\it (lifted) characteristic vectors} (on $T^\ast M$).
When projected on $M$, we obtain  {\it (projected)  characteristic vectors} $\dot x\in TM$ (on $M$)  tangent to {\it (projected) characteristic curves}.

\Note{
If we fix $x_0$ and we consider all (projected) characteristics through it (for any initial characteristic covector $\xi\in C^\ast_p$), that defines a surface $\nu_{x_0}\subset M$, called the {\it wavefront}
(or, in analytic language, the {\it singular support}), which is regular in a neighbourhood of $x_0$ except at $x_0$.
If we start from Maxwell equations, that becomes the light wavefront at an event $x_0$, which, with a bad naming, is also called the {\it light cone in spacetime} (which, of course, is a {\it coinoid}, not a cone. In fact, since, in general, in a Lorentzian manifold, we have no (intrinsic) definition for $\la x$).
Explicitly, a point $x$ in the wavefront is characterized by the property that there exists a projected characteristics $\ga$ connecting $x$ and $x_0$.

One could (correctly) imagine that we are defining the causal structure on spacetime $M$ as a byproduct of field equations, rather than as given by a metric, which is not there on a bare manifold, or in a background free theory, until we fix a solution of field equations. 

The above is the really physical and only way to go.
}

Unlike what we did in the case of strictly hyperbolic systems, for symmetric hyperbolic systems, in general we can only consider a 1-dimensional subspace $V^\ast=\langle\xi\rangle$ in $C^\ast$,
hence the polar $V_\xi = (V^\ast)^\circ$ will be of dimension $m-1$. It contains the characteristic vectors $v$ along with other tangent vectors to the wavefront $\nu_{x_0}$.

\Note{
One can show that if the system is strictly hyperbolic and we have a whole subspace $V^\ast\subset C^\ast$, then, for any characteristic covector $\xi\in V^\ast$, one defines a different characteristic vector $v$, which in turn defines a subspace $V=\langle v\rangle \subset TM$.
However, different characteristic covectors $\xi\in V^\ast$ define parallel characteristic vectors, hence they all define the same subspace $V$, which agrees with what we did above, namely $V=(V^\ast)^\circ$.
}

In a general symmetric hyperbolic system, we can manipulate the equations similarly to what we did for strictly hyperbolic systems, though not exactly alike.
This time, we fix a characteristic covector $\xi= \al dt+ \xi_a dk^a\in C^\ast_p$, and we find the hyperplane of vectors $\langle \xi\rangle^\circ$ tangent to the characteristic surface $\nu_{x_0}$
and within them the characteristic vector $\dot x$ using (\ShowLabel{HanitonianSystem}).
A basis of $\langle \xi\rangle ^\circ$ is in fact
\eq{
v_a = -\xi_a \del_0 + \al \del_a
}
while {the characteristic vector is given by} $\dot x= \al^a v_a$ with 
\eq{
\al^a = \Frac[\del H/ \del \xi_a]
}

In this case, we cannot reduce the system to derivatives in the direction of $\dot x$. However, we still can do that in the direction of a  vector $v$ tangent to the characteristic surface {$\nu_x$}.
The proof goes along similar lines, as
\eqsLabel{
\al w^i \al^\mu_{ij} y^j_\mu + \al  w^i \be_i 
=& \al w^i \al^0_{ij} y^j_0 + \al w^i \al^a_{ij} y^j_a + \al  w^i \be_i =\cr
=& w^i \al^a_{ij}  \(  -\xi_a  y^j_0 + \al y^j_a\)  + \al  w^i \be_i =\cr
=&w^i \al^a_{ij} (  v^\mu _a  y^j_\mu ) + \al  w^i \be_i 
}{ODEq}
since now we have
\eq{
0=w^i \xi_\mu \al^\mu_{ij} = \al w^i \al^0_{ij} + w^i \xi_a \al^a_{ij} 
\quad\then
\al w^i \al^0_{ij} =- w^i \xi_a \al^a_{ij} 
}

{The expression (\ShowLabel{ODEq}) of field equations still depends on the basis $v_a$ chosen for the subspace $\langle \xi \rangle^\circ$.
 One can further change this basis to $(\vec v, \vec u_i)$ where $\vec v$ is the characteristic vector and $\vec u_i$ are adapted the Cauchy surface $S_0\subset M$.
}

{Since the derivatives of fields along $\vec u_i$ can be computed out of the initial conditions, the only evolution vector is the characteristic vector $\vec v$.
As we shall show in Appendix A, that leads to a form of field equations in which equations are ODE along characteristic curves, as we did for strictly hyperbolic systems.
}

That shows that the PDE restricts to a PDE on the characteristic surface $\nu_{x_0}$ {or even an ODE along characteristic curves}.
Again, Cauchy surfaces must be transverse to the characteristic surfaces and this says that, knowing fields in a region $U\subset S_0$, we can predict evolution of fields and their derivatives
within a dependency region which is again determined by characteristic surfaces.
(See Appendix $A$ for some detailed examples.)
For a quasi-linear second order Euler-Lagrange equations in the form
\eqLabel{
\al^{\mu\nu}_{ik} y^k_{\mu\nu} + \be_i=0
}{SecondOrderPDE}
the coefficients can depend on $j^0y=(x^\mu, y^i)$ or on $j^1y=(x^\mu, y^i, y^i_\mu)$. 
Both are always intrinsic, while depending on $(x^\mu)$ only is not intrinsic, unless the configuration bundle is affine or a vector bundle (or has a reduction to $\GL(k)$ as it happens to $\Lor(M)$ which, clearly, is not a vector bundle).
The principal symbol of a second order operator is a (global) map 
\eq{
\si: \pi^\ast(S^0_2 M)\arr V^\ast(C)\otimes V^\ast(C)\otimes A_m(M)
}
which is now a quadratic form on covectors. Accordingly, we set
\eq{
\si_{ik}(\xi) = \al^{\mu\nu}_{ik} \xi_\mu\xi_\nu
}
This can be (equivalently) investigated by defining
characteristic covectors in each family as points of a quadric, thus not a subspace any longer. For example, when the characteristic polynomial has a factor $-(\xi_0)^2+(\xi_1)^2+ (\xi_2)^2$, in dimension $m=3$ one gets characteristic covectors
$\xi=\la(dt+ \cos(\te)dx + \sin(\te)dy)$, each corresponding to a $(m-1)$-plane of characteristic vectors, one for each value of $\te$, namely $v=v^1 (-\cos(\te)\del_t + \del_x) + v^2(-\sin(\te) \del_t +\del_y)$.

Or, alternatively and equivalently, we can recast the second order system on $C$ as a first order system (with a constraint $\del_a y^i_b=\del_b y^i_a$) and treat it as above.
Either ways, when we have a symmetric hyperbolic system, the corresponding Cauchy problem is well posed and we get a solution for any initial condition.
Now go back to a covariant equation, which in view of the hole argument (see the Introduction), splits into a boundary part and a bulk part for the subset of bulk fields. 
Boundary equations constrain allowed initial conditions, the bulk equations define a well posed Cauchy problem for bulk fields, and uniquely determine bulk fields on the base manifold.
Roughly speaking, if we start from allowed initial conditions and we fix gauge fields to parameterize conventions freedom, we obtain covariant fields on $M$ which solve the original equations.
Different choices of gauge fields account for under-determination predicted by the hole argument, boundary equations spoil existence for not-allowed initial conditions (with respect to the original equations, the Cauchy problem is still well posed, but the solution does not solve the original covariant field equations),
accounting for over-determination.

This may seem a bit complicated. However, in fundamental physics, it is rather the scheme all field equations follow, since essentially any fundamental field theory is a gauge-natural theory.

 \section{Setting a Cauchy problem in field theory}
 
 We can set up a framework to discuss  Cauchy problems on a bare manifold $M$.
 For the sake of simplicity, consider a quasi-linear first order problem as above in (\ShowLabel{FirstOrderPDE}), a compact region $\bar D\subset M$ and let us denote by $\del D$ its {\it boundary} and by $D$ its {\it interior}.
 Typical example is an $m$-disk $\bar D$ with a boundary which is topologically a sphere $\del D= S_{m-1}$ 
 (although one sometimes can consider some limit, for example to spatial infinity, which however, needs extra care for mathematical consistency).
 One needs a vector field $\ze$ to describe {\it evolution}, chosen to be compactly supported into $\bar D$ (and non-zero into the interior $D$).
 
 \Note{
 By working within a compact region $\bar D$, which can be selected to be topologically trivial as a disk, one can avoid to deal with global topological constraints on spacetime, 
 and have a better control of asymptotic fields. 
 Moreover, with a partition of unity we can easily produce compactly supported smooth evolution vector fields. 
  }
 
 The evolution vector field $\ze$ defines a flow $\Phi_s$ on $D$ (the vector fields $\ze$ is complete on a compact $\bar D$) 
 and integral curves $\ga:\R \arr D$ which foliate $D$.
 However, foliation is not enough.

\Note{
Consider a linear flow on a torus generated by $X= a\del_\al + b \del_\be$. 
Depending on whether $a/b\in\Q$ or not, the integral curves can be closed or dense.
Of course, when they are dense, they get the wrong induced topology for it to be an embedding. 
Imagine one set initial conditions on a  submanifold $S$ which is dense in $D$: one could then extend fields everywhere simply by continuity, or fail to extend them anyway, without even considering field equations!
}

It is clear, we need a stricter structure which ensures that the integral curves of the evolution field (which we called {\it rest motions}) are submanifolds in $D$.
In particular, we define on $D$ an equivalence relation as $p\sim q$ if they belong to the same integral curve so that $D/\sim$ is the quotient space of rest motions and we require $S=D/\sim$
to be  a manifold, so that we can define a projection $\pi: D\arr S: p\mapsto [q]$ (when $p\sim q$) so that it defines a bundle over $S$ which is called the {\it rest bundle}.

\Note{
The rest bundle allows to discuss Cauchy problems, in which one gives the value of fields on a section $S_0$ of the rest bundle and we can look for a solution in $D$.
Cauchy problems describe evolution and determinism (at least within $D$).
In fact, the evolution $\ze$ is generically transverse to $\del D$, thus integral curves extends by continuity to a map on the boundary. The points $\ga(s)$ on boundary corresponds on the integral curves to $s=\pm \infty$.
That divides the boundary into an initial boundary $\del D_-$  where $\ze$ is pointing inward, and a final boundary $\del D_+$ where where $\ze$ is pointing outward, except some tangent points.

This is especially clear on a disk, for the sake of simplicity in dimension 2.
Let us consider an open disc $D=\{(x, y): x^2+y^2<1\}\subset \R^2$ and the vector field 
$\ze= (1-x^2 - y^2)\del_y$ (on $\bar D$)  which has support in the closure $\bar D$ and it is non-zero and smooth everywhere in $D$.

Integral curves are $\ga_{(x, y)}:\R\arr D: s\mapsto (x, y(s; x, y))$  with
\eq{
y(s; x, y)  = \al \Frac[ \al \sinh(\al s)   +y \cosh(\al s)  /  \al  \cosh(\al s) +y \sinh(\al s)  ]
=  \al \Frac[ \al \tanh(\al s)   +y   /  \al  +y \tanh(\al s)  ]
}
and where we set $\al= \sqrt{1-x^2}$. The latter is real since in $D$ we have $x\in (-1,1)$.
Let us remark that the limits
\eq{
\lim_{s\arr +\infty} y(s; x, y)=  \al = \sqrt{1-x^2}
\qquad
\lim_{s\arr -\infty} y(s; x, y)=  -\al = -\sqrt{1-x^2}
}
show that integral curves have asymptotic limits on the boundary of the disc.
}

Let then $\si:S\arr D$ be a section of the rest bundle and let us denote by $S_0=\si(S)$ its image in $D$. 
It is by construction transverse to the evolution field $\ze$, which is in fact vertical on the rest bundle.
This section can be dragged along the evolution flow $S_s= \Phi_s(S_0)$ to define another foliation of $D$. Hence, we have $D\simeq \R\times S$ and we can define another bundle projecting on the second factor
$p:D\arr \R$, which is called the {\it ADM bundle}.

\Note{
Both the bundles are trivial by two different arguments. For $p:D\arr \R$, 
the ADM bundle is trivial because the base manifold $\R$ is contractible and any bundle on a contractible base is trivial, by a homotopy argument (see \cite{Steenrod}).

For $\pi: D\arr S$, the rest bundle is at the same time a principal bundle for the group $(\R, +)$ and an affine bundle. Being affine, it has global sections, being principal with a global section, it is trivial. 

On $D$, we use coordinates which are fibered along both fibrations, namely $(t, k^a)$ adapted to both rest motions and space submanifolds $S_s$. One can change them as on the rest bundle
\eq{
t'=t'(t, k)
\qquad\qquad
k'^a = k'^a(k)
}
as on the ADM bundle
\eq{
t'=t'(t)
\qquad\qquad
k'^a = k'^a(t, k)
}
or as on the product
\eq{
t'=t'(t)
\qquad\qquad
k'^a = k'^a(k)
}
depending on the physical structure we wish to preserve.
}

In general, the characteristic curves defined by field equations depend on the fields themselves (through the coefficients of the principal symbols).
They do in GR, they necessarily do on a general (not linear) configuration bundle $[\pi:C\arr M]$, they can in any quasi-linear system.

Let us call an {\it evolution bubble} any such a triple $(\bar D, \ze, \si)$ which contains the structure to formulate a Cauchy problem on $D$. 
For each evolution bubble, we can restrict to fields $y:M\arr C$ for which the characteristics are transverse to $S_t$.
By changing the section $\si$ and the evolution field $\ze$, we can cover all sections $y:M\arr C$.

\Note{
If one of the fields is a Lorentzian metric, Einstein equations define characteristics which are lightlike for the field $g$, thus if $S_t$ has to be transverse to light rays, $S_t$ has to be spacelike with respect to $g$.
Thus, we can consider all metrics which makes $S_0$ spacelike and the evolution timelike so that all $S_t$ are transverse. 
The region $D\simeq \R\times S$ is globally hyperbolic (which is necessary to have a well posed Cauchy problem). Of course, on a disk, there is no obstruction to be ``globally'' hyperbolic.    
}

Given a Cauchy bubble $(\bar D, \ze, \si)$, we define the rest bundle $D\arr S$ and we set a Cauchy problem for bulk fields on $D$ with initial conditions on $S_0=\si(S)$ which satisfies boundary equations written on $S_0$.
We aim to use evolution equations to determine bulk fields in $D$ and  rebuild a solution of the covariant equations on $D$.

 We have two issues with this scenario.
Firstly, we need to guarantee that the Cauchy problem in the bulk for the bulk fields is well posed.
This is usually done by resorting to a symmetric hyperbolic system for which Cauchy problems are well posed. 
This is true for smooth initial conditions. For initial conditions in some Sobolev space, one is guaranteed to find solution in another Sobolev space. 
However, we are not interested here in non-smooth solutions.
The condition is sufficient, that is, it is not the only way one can have well-posed Cauchy problems. 
For example, one often can recast a non-symmetric-hyperbolic system into a symmetric-hyperbolic one (e.g.~by adding a condition, see Appendix $A$) and find a unique solution anyway.

Secondly, we have constraints on initial conditions. Of course, constraint equations do not spoil well-posedness. However, we impose them on initial conditions, while we need to guarantee that they are met at any $S_t$ during evolution.
This is a compatibility condition between evolution and constraint equations which has to be met if we want to be able to eventually rebuild a solution of the original covariant equations.
In standard GR, this is guaranteed by Bianchi identities (see \cite{Taylor}, \cite{Gar1}). Further investigation is in order to deal with a general (gauge-)natural theory.

In any event, this analysis accounts for what to do when looking for numerical solutions by discretising the Cauchy problem, find a solution of bulk fields which satisfies both the evolution equations {\it and} the constraint equations,
and finally use bulk fields and gauge fields to define a solution of the covariant field equations.

\section{Pre-quantum states}

Given an evolution bubble $(\bar D, \ze, \si)$, we can pull-back the configuration bundle along the embedding $i:\del D\arr M$. 
A section of this pulled-back bundle represents the value of fields on the boundary, and it is called {\it pre-quantum configuration}.
If it satisfies boundary equations, it is called {\it pre-quantum state}.
Any pre-quantum state contains boundary information about the value of fields (given on $\del D = \del D_-\union \del D_+$) which satisfy boundary equations.

In a quantum perspective, the bulk solution is not expected to be meaningful. 
In mechanics, the bulk equations would determine the trajectory of a particle that we know is not something well defined in a quantum model.
For that reason, a quantum model focuses on boundary equations only. In a sense, pre-quantum states are in correspondence with solutions, they are a representation of the phase space of the system, and we could even say that focusing on the solution space is what Hamiltonian formalism and phase space are made for.

Boundary equations select among pre-quantum configurations the subset of pre-quantum states which are the boundary values of fields such that there exists a classical bulk solution which agrees with those boundary values.
Finding such a bulk solution defines the classical propagator, a solution connecting the initial state on $\del D_-$ to the final state on $\del D_+$.
The boundary equations naturally act on pre-quantum configurations which in fact contain all information available at a quantum level.
From a quantum viewpoint, all a quantum model is required to do is associating a probability amplitude to a boundary state, which means defining the quantum propagator.

In a nutshell, LQG is a non-perturbative, generally covariant, background free quantisation of the boundary equations of the ABI model we discussed in the previous paper (see \cite{LN1}), as we shall show in the next lecture notes
(see \cite{Rovelli1}, \cite{Rovelli2}).

\section{Conclusions and perspectives}

We discussed in a Lagrangian framework how one can formulate Cauchy problems for a generally covariant theory.
That is equivalent to the canonical analysis based on Hamiltonian formalism and it shows how the typical situation in relativistic or gauge theories is quite complicated.
One cannot assume a metric given on spacetime, or confuse covectors and vectors. 
However, the principal symbol allows to classify well posed Cauchy problems in an evolution bubble. 
This is less than solving a global Cauchy problem on $M$ (for which we have a number of important results (see \cite{Choquet}). 
On the other hand, being defined on a compact set, one can avoid most of the topological constraints, in the sense that the evolution bubble can always be chosen to be topologically trivial.
On the other hand, one assumes to know the fields everywhere outside the bubble and to constrain values on the boundary so that it extends inside and
compute the solution inside  (which is a sort of classical control problem).

This clarifies the different attitude between a classical and a quantum perspective, since in the second one the value of fields in the bubble are undefined from a quantum physical point of view.
The standard canonical analysis based on Hamiltonian formalism offers more geometric structures to work, and often it is computationally easier.
On the other hand, it is non-canonical in field theory as well as it partially spoils manifest general covariance, which needs to be restored {\it  a posteriori}.

Both Lagrangian and Hamiltonian formalisms are important and each gives us something. 
Both are concerned with investigating the space of solutions of field equations which is the core of Hamiltonian formalism regardless its technical setting.
The next lecture notes will be devoted to analysing the ABI model along the lines presented here.

\section*{Appendix: Examples of Characteristics}

We now give some simple examples of application of the theory of characteristics illustrated above, initially focusing, for simplicity,
on scalar fields for the bidimensional case $M=\R^2$. This will illustrate the link of our approach with the classical technique 
for solving semilinear and quasilinear partial differential equations by means of the method of characteristics. In the sequel
of this section we often write $\bfx=(x_0,x_1)=(t,x)$, $(\xi_0,\xi_1)=(\tau,\xi)$, or, alternatively, $\bfx=(x_0,x_1)=(x,y)$, $(\xi_0,\xi_1)=(\xi,\eta)$.

\subsection{A semilinear example}\label{exmp:exa}
	%
	We first consider a very elementary example of scalar field $u$ on $M=\R^2_{(t,x)}$, that is
\eqLabel{
		u_0-u_1=\beta(\cdot,u)
}{eq:exmpl_1}
	which corresponds to the semilinear PDE
	\eqLabel{
			\partial_t u(t,x)-\partial_x u(t,x)=\beta(t,x,u(t,x))
	}{eq:smlinpde_1}
	Here we have $\sigma=\xi_0-\xi_1$, we need to solve the (unidimensional) equation $w^i\sigma^\mu_{ij}\xi_\mu=w(\xi_0-\xi_1)=0$.
	We obtain the characteristic family $w=\partial_u$, and $\xi=\alpha(dt+dx)$, $\alpha\in\R$.
	The dual characteristic cone $\cC^\alpha$ is spanned by $\xi$, that is, 
	\eq{
	\Align{
		(\langle\xi\rangle)^\circ=&\{v=a(\partial_0-\partial_1),a\in\R\}=\cr
		=&\left\{v=a\left(\dfrac{\partial}{\partial t}-\dfrac{\partial}{\partial x}\right),a\in\R\right\}\Leftrightarrow \xi(v)=0
		}
	}
	The characteristics are 
	\eqLabel{
		\begin{cases}
			t=t_0+s
			\\
			x=x_0-s, & s\in\R
		\end{cases}
	}{eq:charc_1}
	Indeed, given the Hamiltonian $H=\det\sigma=\xi_0-\xi_1=\tau-\xi$, it follows
	\eq{
		\begin{cases}
			\dot{t}=\dfrac{\partial H}{\partial \tau}(t,x,\tau,\xi)=1
			\\
			\dot{x}=\dfrac{\partial H}{\partial \xi}(t,x,\tau,\xi)=-1\rule{0mm}{7mm}
		\end{cases}
	}
	which implies \eqref{eq:charc_1}. Setting $U(s)=u(t_0+s,x_0-s)$, it immediately follows 
	\eq{
		\dot{U}(s)=(\partial_t u)(t_0+s,x_0-s)-(\partial_x u)(t_0+s,x_0-s)
	}
	and \eqref{eq:smlinpde_1} becomes the ODE
	\eq{
		\dot{U}(s)=\beta(t_0+s,x_0-s,U(s))
	}
	as claimed in Section \ref{S2}. The \textit{lifted characteristics} are the integral curves of the vector field 
	\eq{
		X=\displaystyle\Latexfrac{\partial}{\partial t}-\Latexfrac{\partial}{\partial x}+\beta(t,x,u)\Latexfrac{\partial}{\partial u},
	}
	namely,
	\eq{
		dt=-dx=\Latexfrac{du}{\beta(t,x,u)}\Leftrightarrow
		\begin{cases}
			t=t_0+s,
			\\
			x=x_0-s,
			\\
			\dot{u}(s)=\beta(t_0+s,x_0-s,u(s)), &s\in\R.
		\end{cases}
	}
	Recall that the latter naturally arise in solving \eqref{eq:smlinpde_1}, by looking for an auxiliary function
	$v=v(t,x,u)$ such that $v(t,x,u(t,x))\equiv 0$ and 
	$\partial_u v(t,x,u)\not\equiv0$.
	Then, by the Dini theorem, 
	\eq{
		\partial_t u(t,x)=-(\partial_t v)(t,x,u(t,x))/(\partial_u v)(t,x,u(t,x))
	}
	and
	\eq{
		\partial_x u(t,x)=-(\partial_x v)(t,x,u(t,x))/(\partial_u v)(t,x,u(t,x)),
	}
	which, substituted in \eqref{eq:smlinpde_1}, give
	\eq{
	\Align{
	&	-\Latexfrac{\partial_t v(t,x,u)}{\partial_u v(t,x,u)}+\Latexfrac{\partial_x v(t,x,u)}{\partial_u v(t,x,u)}=\beta(t,x,u)\cr
	&	\Leftrightarrow
		\partial_t v(t,x,u) - \partial_x v(t,x,u)+\beta(t,x,u)\partial_u v(t,x,u) = 0.
	}
	}
	%
%
\subsection{A point-dependant semilinear example}\label{exmp:exb}
	%
	Similarly to Example \ref{exmp:exa}, consider, always in the scalar case,
	\eqLabel{
		u_0(\bfx)-x_1u_1(\bfx)=\beta(\bfx,u(\bfx))
	}{eq:exmpl_2}
	which corresponds to the semilinear PDE
	\eqLabel{
		\partial_t u(t,x)-x\partial_x u(t,x)=\beta(x,u(x))
	}{eq:smlinpde_2}
	Here we have $\sigma=\tau-x\xi$, 
	$w=\partial_u$, $\xi=\alpha(xdt+dx)$, $\alpha\in\R$, 
	is the characteristic family, and $v=\dfrac{\partial}{\partial t}-x\dfrac{\partial}{\partial x}$ is characteristic.
	The characteristics are 
	\eqLabel{
		\begin{cases}
			\dot{t}=1
			\\
			\dot{x}=-x
		\end{cases}
		\Leftrightarrow
		\begin{cases}
			t=t_0+s
			\\
			x=x_0e^{-s}, & s\in\R
		\end{cases}
	}{eq:charc_2}
	Indeed, given the Hamiltonian $H=\det\sigma=\xi_0-x_1\xi_1=\tau-x\xi$, it follows
	\eq{
		\begin{cases}
			\dot{t}=\dfrac{\partial H}{\partial \tau}(t,x,\tau,\xi)=1
			\\
			\dot{x}=\dfrac{\partial H}{\partial \xi}(t,x,\tau,\xi)=-x\rule{0mm}{7mm}
		\end{cases}
	}
	which implies \eqref{eq:charc_2}. Setting $U(s)=u(t_0+s,x_0e^{-s})$ we of course have
	\eq{
		\dot{U}(s)=(\partial_t u)(t_0+s,x_0e^{-s})-x_0e^{-s}(\partial_x u)(t_0+s,x_0e^{-s})
	}
	and \eqref{eq:smlinpde_2} becomes the ODE
	\eq{
		\dot{U}(s)=\beta(t_0+s,x_0e^{-s},U(s))
	}
	The lifted characteristics are the integral curves of the vector field 
	\eq{
		X=\displaystyle\Latexfrac{\partial}{\partial t}-x\Latexfrac{\partial}{\partial x}+\beta(t,x,u)\Latexfrac{\partial}{\partial u}
	}
	namely,
	\eq{
		dt=-\Latexfrac{dx}{x}=\Latexfrac{du}{\beta(t,x,u)}\Leftrightarrow
		\begin{cases}
			t=t_0+s
			\\
			x=x_0e^{-s}
			\\
			\dot{u}(s)=\beta(t_0+s,x_0e^{-s},u(s)), &s\in\R
		\end{cases}
	}
	As in Example \ref{exmp:exa}, solving \eqref{eq:smlinpde_2} by looking for $v=v(t,x,u)$ such that $v(t,x,u(t,x))\equiv 0$ and 
	$\partial_u v(t,x,u)\not\equiv0$, we find
	\eq{
	\Align{
	&	-\Latexfrac{\partial_t v(t,x,u)}{\partial_u v(t,x,u)}+x\Latexfrac{\partial_x v(t,x,u)}{\partial_u v(t,x,u)}=\beta(t,x,u)\cr
	&	\Leftrightarrow
		\partial_t v(t,x,u) - x\partial_x v(t,x,u)+\beta(t,x,u)\partial_u v(t,x,u) = 0.
	}
	}
	%
%
\subsection{A quasilinear example}\label{exmp:exc}
	%
	Consider
	\eqLabel{
		u_0-u\,u_1=\beta(\cdot,u)
	}{eq:exmpl_3}
	which corresponds to the quasilinear PDE
	\eqLabel{
		\partial_t u(t,x)-u(t,x)\partial_x u(t,x)=\beta(t,x,u(t,x))
	}{eq:smlinpde_3}
	Here we have $\sigma=\xi_0-u\xi_1$, 
	$w=\partial_u$, $\xi=\alpha(udt+dx)$, $\alpha\in\R$,
	is the characteristic family, and $v=\dfrac{\partial}{\partial t}-u\dfrac{\partial}{\partial x}$ is characteristic.
	The lifted characteristics are the integral curves of the vector field 
	\eq{
		X=\displaystyle\Latexfrac{\partial}{\partial t}-u\Latexfrac{\partial}{\partial x}+\beta(t,x,u)\Latexfrac{\partial}{\partial u}
	}
	namely,
	\eq{
		dt=-\Latexfrac{dx}{x}=\Latexfrac{du}{\beta(t,x,u)}\Leftrightarrow
		\begin{cases}
			\dot{t}=1
			\\
			\dot{x}=-u
			\\
			\dot{u}=\beta(t,x,u)
		\end{cases}
	}
	associated with the (linearized) PDE
	\eq{
	\Align{
	&	\partial_t v(t,x,u) - u\partial_x v(t,x,u)+\beta(t,x,u)\partial_u v(t,x,u) = 0 \cr 
	&	v(t,x,u(t,x))\equiv0, \partial_uv(t,x,u)\not\equiv 0
	}
	}
	%
%
\subsection{A point-dependent quasilinear example}\label{exmp:exd}
	%
	Let us go a little further ahead with respect to the previous examples, considering
	\eqLabel{ 
		x_1\,u(\bfx)\,u_0(\bfx)+x_0\,u(\bfx)\,u_1(\bfx)=[u(\bfx)]^2
	}{eq:exmpl_4}
	which corresponds to the quasilinear PDE
	\eqLabel{ 
		y\,u(x,y)\,\partial_x u(x,y)+x\,u(x,y)\,\partial_y u(x,y)=[u(x,y)]^2
	}{eq:smlinpde_4}
	Here $\sigma=u(x_1\xi_0+x_0\xi_1)=u(x\xi+y\eta)$
	$w=\partial_u$, $\xi=\alpha u(xdx-ydy)$, $\alpha\in\R$
	is the characteristic family, and $v=y\dfrac{\partial}{\partial x}+x\dfrac{\partial}{\partial y}$ is characteristic.
	The lifted characteristics are the integral curves of the vector field 
	\eq{
		X=\displaystyle yu\Latexfrac{\partial}{\partial x}+xu\Latexfrac{\partial}{\partial y}+u^2\Latexfrac{\partial}{\partial u}
	}
	namely,
	\eq{
		\Latexfrac{dx}{uy}=-\Latexfrac{dy}{ux}=\Latexfrac{du}{u^2}\Leftrightarrow
		\begin{cases}
			\dot{x}=uy
			\\
			\dot{y}=ux
			\\
			\dot{u}=u^2
		\end{cases}
	}
	It follows
	\eq{
	\Align{
		&
		\begin{cases}
			u(s)=\dfrac{1}{c_1-s}
			\\
			x^\prime(u)=\dfrac{y(u)}{u}\rule{0mm}{7mm}
			\\
			y^\prime(u)=\dfrac{x(u)}{u}\rule{0mm}{7mm}
		\end{cases}		
		\Rightarrow
		\begin{cases}
			u(s)=\dfrac{1}{c_1-s}
			\\
			u^2x^{\prime\prime}(u)+ux^\prime(u)-x(u)=0\rule{0mm}{5mm}		
			\\
			u^2y^{\prime\prime}(u)+uy^\prime(u)-y(u)=0
		\end{cases}	
		\Rightarrow
		\begin{cases}
			u(s)=\dfrac{1}{c_1-s}
			\\
			x(u)=c_2 u +\dfrac{c_3}{u}	\rule{0mm}{7mm}		
			\\
			y(u)=c_2 u - \dfrac{c_3}{u}\rule{0mm}{7mm}
		\end{cases}
		\\	
		\Rightarrow &
		\begin{cases}
			u(s)=\dfrac{u_0}{1-su_0}
			\\
			x(s)=\dfrac{x_0-y_0}{2} (1-su_0) +\dfrac{x_0+y_0}{2}\dfrac{1}{1-su_0}\rule{0mm}{7mm}		
			\\
			y(s)=\dfrac{y_0-x_0}{2} (1-su_0) +\dfrac{x_0+y_0}{2}\dfrac{1}{1-su_0}\rule{0mm}{7mm}
		\end{cases}	
	}
	}
	For the sake of completeness, let us now solve \eqref{eq:smlinpde_4}. As above, we first look for $v=v(t,x,u)$ such that $v(x,y,u(x,y))\equiv 0$ and 
	$\partial_u v(x,y,u)\not\equiv0$, leading us to the linear PDE
	\eqLabel{
		yu\partial_x v(x,y,u) + xu\partial_y v(x,y,u)+u^2\partial_u v(x,y,u) = 0
	}{eq:linearized}
	By the above computations, we see that the family of characteristics can be equivalently written as
	\eq{
		\begin{cases}
			\dfrac{x+y}{u}=k_1
			\\
			(x-y)u=k_2\rule{0mm}{5mm}
		\end{cases}
	}
	for constants $k_1,k_2\in\R$. Then, the solutions of \eqref{eq:linearized} is given by
	\eq{
		v(x,y,u(x,y))=G\left(\dfrac{x+y}{u(x,y)},(x-y)u(x,y)\right)\!
	}
	for an arbitrary function $G\in C^1$. For instance, choosing $G(t_1,t_2)=t_1-t_2$, we obtain $u(x,y)=\sqrt{\dfrac{x+y}{x-y}}$.
	The choice $G(t_1,t_2)=(t_1 -1)t_2$ gives $u(x,y)=x+y$, while choosing $G(t_1,t_2)= t_1 t_2 - \sin t_2$ we find 
	$u(x,y)= \dfrac{\mathrm{arcsin}\,(x^2-y^2)}{x-y}$.
	%

Let us then present some examples of quasilinear systems of different kind and characteristics to analyse them.

\subsection{A strictly hyperbolic system}

Let us consider a quasi-linear (actually, semi-linear in this case), first order system
\eq{
 \Cases{
	u_t+u_x+4u_y - v_x-v_y + b_1(u, v) =0	\cr
	-u_x -u_y +v_t -\frac[1/2] v_x+\frac[5/2] v_y+ b_2(u, v) =0	\cr
 }
}
Its principal symbol reads as
\eq{
\si_{ij}=\(\Matrix{
\xi_0+\xi_1+4\xi_2	&	-\xi_1-\xi_2	\cr
-\xi_1-\xi_2		&	\xi_0	-\frac[1/2]\xi_1+ \frac[5/2]\xi_2	\cr
}\)
}
which is symmetric.
The determinant of the principal symbol is
\eq{
Q(b;\xi)=\frac[1/2](2\xi_0 + 3\xi_1 +9  \xi_2)(\xi_0 - \xi_1 +2\xi_2)
}

Then, we have two subspaces of characteristic covectors, namely  
\eq{
 \Align{
 \xi^1 =& \al (-\frac[3/2] dt +dx) +\be (-\frac[9/2]dt + dy)=  -( \frac[3/2] \al +\frac[9/2] \be ) dt +\al dx  + \be dy \cr
 \xi^2 =& \al (dt +dx) +\be (-2 dt+dy)=  (\al  -2 \be) dt  +\al dx+\be dy
 }
}
as well as the corresponding characteristic vectors
\eq{
v_1= v^0 (\del_0 +\frac[3/2]\del_1 + \frac[9/2]\del_2)
\qquad\qquad
v_2=v^0 (\del_0 -\del_1 +2\del_2)
}
There is a basis $(w_1, w_2)$ in $V_b(B)$, given by
\eq{
w_1=  2\del_u - \del_v
\qquad
w_2= \del_u+2 \del_v
}
We consider the linear combinations of field equations 
\eq{
\Cases{
	2u_t -v_t 
+\frac[3/2](2u_x   -  v_x)
+\frac[9/2](2u_y- v_y)
+ 2b_1(u, v) - b_2(u, v)
 =0	\cr	
u_t +2v_t
-(u_x   +2 v_x )
+2(u_y+2 v_y)
+ b_1(u, v) + 2b_2(u, v) 
	=0	\cr
}
}
Accordingly, we define the new fields
\eq{
U= 2u-v
\qquad\qquad
V=u+2v
}
so that the original system is recasted into the form
\eqLabel{
\Cases{
	U_t 
+\frac[3/2] U_x  
+\frac[9/2]U_y
+ 2b_1(u, v) - b_2(u, v)
 =0	\cr	
V_t 
-V_x  
+2V_y
+ b_1(u, v) + 2b_2(u, v) 
	=0\cr
}\qquad\then
\Cases{
	 \dot U = \xi(U, V) 	\cr
	\dot V = \ze(U, V)	\cr
}
}{ODE}

{We can obtain the same result in matrix form. The original system can be written as 
\eq{
\(\Matrix{
\del_0 + \del_x+ 4 \del_x	&	-\del_x-\del_y \cr
-\del_x-\del_y 			&	\del_0-\frac[1/2]\del_1+\frac[5/2]\del_2
}\)\(\Matrix{
u\cr 
v
}\)= \(\Matrix{
-b_1\cr
-b_2
}\)
}
By using the basis $w_i$ of vertical vectors we can define the transformation matrix
\eq{
P= \(\Matrix{
	2	&	1	\cr
	-1	&	2	\cr
}\)
\qquad\then\quad
\bar P=\frac[1/5] \(\Matrix{
	2	&	-1	\cr
	1	&	2	\cr
}\)
}
Then we define new fields 
\eq{
\(\Matrix{
	\hat U	\cr
	\hat V	\cr
}\)= \frac[1/5]\(\Matrix{
	2	&	-1	\cr
	1	&	2	\cr
}\) \(\Matrix{
	u	\cr
	v	\cr
}\)
\qquad\Leftrightarrow\quad
\(\Matrix{
	u	\cr
	v	\cr
}\)=\(\Matrix{
	2	&	1	\cr
	-1	&	2	\cr
}\) \(\Matrix{
	U	\cr
	V	\cr
}\)
}
which agree with $(U, V)$ above up to a constant factor.
}

{
Finally, we can rewrite the system in the new basis as $({}^t P \si P)( \bar P {u \choose v}) = {}^t  P {-b_1 \choose -b_2}$.
We have
\eq{
{}^t P \si P=5 \(\Matrix{
\del_0+\frac[3/2]\del_1+\frac[9/2]\del_2	&	0 \cr
0	&	\del_0-\del_1+2\del_2\cr
}\)
}
which is, in fact, diagonal and the diagonal elements are derivatives along the characteristic vectors.
}

\subsection{A symmetric non-strictly hyperbolic system}

Let us consider the first order, quasi-linear operator, which in fact is symmetric hyperbolic, given by
\eqLabel{
\Cases{
\del_0 u + \del_1u + \del_2 v =b_1	\cr
 \del_2 u +  \del_0 v - \del_1 v=b_2	\cr
}
}{ExPDE}
The system (\ShowLabel{ExPDE}) is translated into a matrix of homogeneous polynomials
\eq{
\(\Matrix{
\xi_0	+ \xi_1	&	\xi_2\cr
\xi_2			&	\xi_0-\xi_1\cr
}\)
}
The dual characteristic cone is
\eq{
(\xi_0)^2 -(\xi_1)^2-(\xi_2)^2=0
\qquad\then
\xi_0 = e_0 \De
}
where we set $e_0=\pm1$ and $\De^2={(\xi_1)^2+  (\xi_2)^2}$.

We choose a characteristic covector $\xi= dt + \cos\te\>dx + \sin\te\> dy \in \calC^\ast$ so that we can associate to it a hyperplane $\langle\xi\rangle^\circ=\langle \vec e_1, \vec e_2\rangle$ of vectors
\eq{
\vec v= v^1 (-\cos\te\del_0 + \del_1) + v^2 (-\sin\te\del_0 + \del_2)
= v^1 \vec e_1 + v^2 \vec e_2
} 
In $\langle\xi\rangle^\circ$ we have the characteristic vector 
\eq{
\dot x = \del_0-\cos\te \del_1-\sin\te \del_2 = -\cos\te \vec e_1 -\sin\te \vec e_2
}
We can also complete the basis of $\langle\xi\rangle^\circ = \langle e_1, e_2\rangle= \langle \vec v, \vec u\rangle$, by choosing another vector
\eq{
\vec u = \sin\te\vec e_1 - \cos\te\vec e_2
}
adapted to the foliation, i.e.~tangent to the Cauchy surface, in this case $S_0:(t=0)$.
Finally, we have a vertical vector
\eq{
w\propto -\sin\te \Frac[\del/\del u] + (1+\cos\te) \Frac[\del/\del v] 
\propto  (1-\cos\te) \Frac[\del/\del u] - \sin\te \Frac[\del/\del v] 
}
in the space of vertical vectors of the configuration bundle.
In this way we have a characteristic family for each value of $\te$.

We can fix two characteristic families, e.g.~let us fix $\te_+= 0$ and $\te_-= \frac[\pi/2]$, so that we have
\eqs{
&
\dot x_+ =\del_0-\del_1
\quad
\vec u_+ = -\del_2 		
\quad
w_+ = 2\Frac[\del/\del v]
\quad\qquad
\vec e_1= - \dot x_+
\quad
\vec e_2= -\vec u_+
\cr
&			
\dot x_- = \del_0-\del_2
\quad
\vec u_- = \del_1	 		
\quad
w_- = -\Frac[\del/\del u] + \Frac[\del/\del v]
\quad\quad
\vec \ep_1= \vec u_-
\quad
\vec \ep_2= -\vec v_-\cr
}

We can use the basis $(w_+, w_-)$ to write the equations. We have the relevant transformation matrices
\eq{
P= \(\Matrix{
0	&	-1	\cr
2	&	1	\cr
}\)
\qquad\then\quad
{}^tP= \(\Matrix{
0	&	2	\cr
-1	&	1	\cr
}\)
\qquad
\bar P= \Frac[1/2]\(\Matrix{
1	&	1	\cr
-2	&	0	\cr
}\)
}
and we can recast the original differential operator $({}^tP \si P) (\bar P {u\choose v}) =  {}^tP {b_1\choose b_2} =  {B_1\choose B_2}$.
We obtain 
\eq{
\si' =  2\(\Matrix{
2 \dot x_+		& \dot x_- -u_-		\cr
\dot x_+ +u_+	&	\dot x_-	\cr
}\)
\qquad\qquad
\(\Matrix{
B_1		\cr
B_2	\cr
}\)=
\(\Matrix{
	2b_2	\cr
	b_2-b_1\cr
}\)
}
for the new fields 
\eq{
\(\Matrix{
U		\cr
V	\cr
}\)= \Frac[1/2]\(\Matrix{
1	&1		\cr
-1	&	0	\cr
}\)\(\Matrix{
u	\cr
v	\cr
}\)
=\(\Matrix{
\frac[u+v/2]	\cr
-u	\cr
}\)
}

This is still written in terms of the basis $(\vec e_1, \vec e_2)$. We can change that basis to an adapted basis. 
This corresponds to perform linear combinations of the equations to obtain
\eqs{
\(\Matrix{
1		&	-1 \cr
-1		&	2\cr
}\)& \(\Matrix{
2 \dot x_+(U)	+ \dot x_-(V) -u_-(V)		\cr
\dot x_+(U) +u_+(U)	+ \dot x_-(V)	\cr
}\) =\cr
=&\(\Matrix{
 \dot x_+(U)	 -u_-(V)	-u_+(U)		\cr
 \dot x_-(V)      + u_-(V)	+ 2u_+(U)	\cr
}\) 
= \Frac[1/2]\(\Matrix{
	b_2 +b_1	\cr
	- b_1\cr
}\)
}

Hence, we have eventually the system in the form
\eq{
\Cases{
 \dot x_+(U)	=  u_-(V)	+u_+(U)  + \Frac[1/2](b_2 +b_1)  \cr
  \dot x_-(V)       =- u_-(V)	- 2u_+(U)  - b_1
 }
}

The derivatives along the Cauchy surface (as $ u_-(V)$ or $u_+(U)$) are known when the initial conditions are known.
The derivatives along characteristic vectors (as $ \dot x_+(U)$ or $\dot x_-(V)$) contain information about the evolution of fields.

Accordingly, if we compute the first equation along the characteristic curve of the first family (as well as the second equation along the characteristic curve of the second family),
they become ODEs (yet depending on both fields $(U, V)$) for the functions
\eq{
U(s)= U\circ \ga_+(s)
\qquad\qquad
V(s)= V\circ \ga_-(s)
}

That is enough to have that if the characteristic curves intersect the Cauchy surface more than once, that alone spoils the possibility of freely given initial conditions.
The initial conditions at the different intersections are constrained by the field equations restricted along the characteristic curves.

\subsection{Characteristics for standard metric GR}

Vacuum Einstein equations can be recast into a quasi-linear, second order, system of PDEs for the metric $g$ on the spacetime $M$. They read as $R_{\al\be}=0$, i.e.~as
\eq{
\frac[1/2]( g^{\tau\te}g_{\al(\rho }g_{\si)\be} - g_{\al(\rho} \de^\tau_{\si)} \de^\te_{\be}
- g_{\be(\rho} \de^\tau_{\si)} \de^\te_{\al} +g_{\rho\si} \de^\tau_\al \de^\te_\be) \del_{\tau\te} g^{\rho\si} = b_{\al\be}(j^1g)
} 
In dimension $m=\dim(M)=4$, these are 10 equations, since they are symmetric with respect to indices $(\al\be)$.
The principal symbol of this system is
\eq{
[\si(\xi)]_{(\al\be)(\rho\si)} = \frac[1/2]( g^{\tau\te}g_{\al(\rho }g_{\si)\be} - g_{\al(\rho} \de^\tau_{\si)} \de^\te_{\be}
- g_{\be(\rho} \de^\tau_{\si)} \de^\te_{\al} +g_{\rho\si} \de^\tau_\al \de^\te_\be)\xi_\tau\xi_\te
} 
which is a $10\times 10$ matrix. At any point $x\in M$, we can always choose coordinates in which $g_{\mu\nu}(x)= \eta_{\mu\nu}$ so that, by ordering the indices pairs as 
\eq{
(0,0)\ (0,1)\ (1,1)\ (2,0)\ (2,1)\ (2,2)\ (3,0)\ (3,1)\ (3,2)\ (3,3)
} 
we get a matrix $[\si(\xi)]_{(\al\be)(\rho\si)}$ given by 
\eq{\footnotesize
\Frac[1/2]\(\Matrix{
\xi_{123}^2	&\xi_{01}				&\xi_0^2		&\xi_{02}			&0				&\xi_0^2	&\xi_{03}			&0				&0				&\xi_0^2 		\cr
0			&-\frac[\xi_{23}^2/2]		&0			&\frac[\xi_{12}/2]	&-\frac[\xi_{02}/2]	&\xi_{01}	&\frac[\xi_{13}/2]	&-\frac[\xi_{03}/2]	&0				&\xi_{01}		\cr
-\xi_1^2		&-\xi_{01}				&\xi_{023}^2	&0				&-\xi_{12}			&\xi_1^2	&0				&-\xi_{13}			&0				&\xi_1^2		\cr
0			& \frac[\xi_{12}/2]		&\xi_{02}		&-\frac[\xi_{13}^2/2]	&-\frac[\xi_{01}/2]	&0		&\frac[\xi_{23}/2]	&0				&-\frac[\xi_{03}/2]	&\xi_{02}		\cr
-\xi_{12}		&-\frac[\xi_{02}/2]		&0			&-\frac[\xi_{01}/2]	&\frac[\xi_{03}^2/2]	&0		&0				&-\frac[\xi_{23}/2]	&-\frac[\xi_{13}/2]	&\xi_{12}		\cr
-\xi_2^2		&0					&\xi_2^2		&-\xi_{02}			&-\xi_{12}\			&\xi_{013}^2	&0			&0				&-\xi_{23}			&\xi_2^2		\cr
0			&\frac[\xi_{13}/2]		&\xi_{03}		&\frac[\xi_{23}/2]	&0				&\xi_{03}	&-\frac[\xi_{12}^2/2]	&-\frac[\xi_{01}/2]	&-\frac[\xi_{02}/2]	&0			\cr
-\xi_{13}		&-\frac[\xi_{03}/2]		&0			&0				&-\frac[\xi_{23}/2]	&\xi_{13}	&-\frac[\xi_{01}/2]	&\frac[\xi_{02}^2/2]	&-\frac[\xi_{12}/2]	&0			\cr
-\xi_{23}		&0					&\xi_{23}		&-\frac[\xi_{03}/2]	&-\frac[\xi_{13}/2]	&0		&-\frac[\xi_{02}/2]	&-\frac[\xi_{12}/2]	&\frac[\xi_{01}^2/2]	&0			\cr
-\xi_3^2		&0					&\xi_3^2		&0				&0				&\xi_3^2	&-\xi_{03}			&-\xi_{13}			&-\xi_{23}			&\xi_{012}^2	\cr
}\)
}
where we set $\xi_{\al\be}:= \xi_\al\xi_\be$,
$\xi^2_{ij}:= \xi^2_i + \xi^2_j$,
$\xi^2_{0j}:= -\xi^2_0 + \xi^2_j$,
$\xi^2_{ijk}:= \xi^2_i + \xi^2_j+ \xi^2_k$,
$\xi^2_{0ij}:= -\xi^2_0 + \xi^2_i + \xi^2_j$,
 for short. 
 This matrix is rank 6, showing that 4 equations cannot be used for a Cauchy problem on $M$.
 This is not surprising, since the covariant system is under-determined by the hole argument.
 
 If we look for directions in spacetime on which the value of some field is determined by equations we find that rank drops (to 3) along all $g$-lightlike covectors $\xi$, meaning that if we know the value of fields along a characteristic ray in spacetime, that determines the value of some field along the whole line.
This means that if we set initial conditions on a surface $S$ and a characteristic  ray intersects the surface more than once, we are not free to choose initial conditions freely on $S$.
Of course, this would mean we need $S$ to be spacelike (and spacetime to be causal, so that we know light rays intersect $S$ only once).
However, we have no metric defined on spacetime yet. We still have to write field equations, choose initial conditions, determine a solution before we have a metric.
This is what we mean by saying that GR is background free or that we are doing a model of gravity in differential topology.

Hence, there is only a way to describe the situation (which, by the way, is exactly what happens with adapted fields, later in \cite{LN3}): 
we choose a surface $S$ and, by fixing it, we can consider only metric fields which make it spacelike. If we want to consider all metrics, we need to change initial surface. Any metric allows a surface which is spacelike, thus we are sure that we are in fact considering any Lorentzian metrics (more or less as it happens by changing coordinates on a manifold), but by choosing an initial surface fixes a subset of metrics we are able to discuss on that surface.

\Note{
This does not mean we have preferred directions on spacetime. Any direction is lightlike for some metric (again by an argument which resorts to compactly supported diffeomorphisms as the hole argument).
This is again a consequence of general covariance (as the hole argument). We {\it will eventually} have preferred directions once we shall have a metric. 
The preferred direction is not on spacetime, it is defined by any specific gravitational field: it is not there before we fix a specific gravitational field.

This is analogous to proper time. Proper time is undefined along a worldline, it comes with the gravitational field. (Proper) time is not there on spacetime, namely in GR, it comes with the gravitational field, i.e.~{\it we} define it using the gravitational field.
}

Since we are left with 6 independent equations, we can hope to split fields into 4 gauge and 6 bulk fields adapted to the foliation, so that equations only determine the bulk fields evolution along the foliation.
We have space leaves $S_t\simeq S\subset M$ and the spacetime metric $g$ induces a metric $\ga$ on each leaf, the lapse $N$ and the shift $\be^i$.
The transformation $(g_{\mu\nu})\mapsto (N, \be^i, \ga^{ij})$ is one-to-one and we can write (vacuum) Einstein equations in terms of the new fields as
\eq{
\Cases{
\calH:=\del_0 K - \be^k D_k K- D_i D^i N +NK^{ij}K_{ij}=0\cr
\calM_i := \(\ga^{lj}\de_i^k - \ga^{kj}\de_i^l\) D_l K_{kj}=0\cr
\(\ga_{il} \del_0 K^l{}_j -   \be^k D_k  K_{ij} + D_k \be_i K^k{}_j -  D_j\be^k K_{ik} - D_iD_j N\)+\cr
\qquad +N\>{}^3\<R_{ij} +NKK_{ij}=0\cr
}
}
where $D_i$ denotes the covariant derivative on $S$ with respect to the induced metric $\ga$ and we set $K_{ij} := N^{-1}\( \frac[1/2]\del_0\ \ga_{ij} -\ga_{k(j} D_{i)} \be^k\)$ for the {\it extrinsic curvature} of the leaf in the spacetime.
Of course, we mean that Latin indices are raised and lowered by the metric $\ga$ (see \cite{Gourgoulhon}).

\Note{
These {\it are} Einstein equations just written with respect to adapted fields.
We just used some facts about submanifolds and foliations in manifolds, which are, in fact, what we originally called {\it differential geometry} (or, if you prefer, its original motivation).
Adapted fields are a mean to expose further properties of the equations, which by the way include physical facts which are traditionally extracted by using Hamiltonian formalism, although they are properties of the equations, not of the formalism one uses to make them manifest. 
}

In this form, Einstein equations are written as a field theory on $S$ aiming to determine families of time-dependent solutions $(N(t, k), \be^k(t, k), \ga^{ij}(t, k)) \mapsto g_{\mu\nu}(t, k)$
which are {\it also} metrics on spacetime, just written in coordinates adapted to the foliations.

\Note{
Notice that if we have a solution $\ga$, we build a spacetime metric $g$ for which, {\it by construction}, the surface $S$ is spacelike. 
As we said, by fixing $S$, we can consider only these metrics, general metrics are obtained by changing $S$.

}

We notice that equations do not contain second time-derivatives of the fields $N$ and $\be^k$.
They are not able to determine the evolution of the lapse $N$ and shift $\be^k$, which is a good thing, since we chose them arbitrarily when we fixed the evolution field and we derived them from it and the foliation. 
As a matter of fact, the lapse and shift fields parameterize the choice of the observers for a foliation and a (transverse) time-direction for evolution, namely, they are gauge fields.
They correspond to the four covariant equations which we found to be redundant: when we drop them, we need to drop as many gauge fields and focus on the evolution of the other bulk fields, in this case $\ga^{ij}$.   
The first equation  $\calH=0$ is called the {\it Wheeler-deWitt equation} or the {\it Hamiltonian constraint}, while the other three boundary equations $\calM_i=0$ are called the {\it momentum constraints}. 
They express arbitrary choices of foliation and coordinates on $S$, respectively.

The other 6 equations are called {\it evolution equations} and they define a Cauchy problem for the induced metric $\ga^{ij}$.
Hence, we set initial conditions $(\ga^{ij}(0, k), \del_0 \ga^{ij}(0, k))$ on $S$ and we get back one and only one solution $\ga^{ij}(t, k)$.
Once we have a solution $\ga^{ij}$, that is not it. We are discussing the original covariant Einstein equations, not the spatial ones. 
The statement is that if all 10 equations are satisfied, then $\ga$ together with $N$ and $\be$ define a solution $g$ of Einstein equations (see
\cite{Gar1}, \cite{Gar2} for details).

If we fix gauge fields at will,  the constraint equations are what the name suggest: they constrain initial conditions for the Cauchy problem for $\ga$.
This is why the covariant equations are {\it also} over-determined. 
When we fix gauge fields and we consider $\ga^{ij}$ as the dynamical field, we have 10 equations for 6 fields. 
The symbol of this system is
\eqLabel{
\si(\xi) = \Frac[1/2]\(\Matrix{
2\xi^2_{23}	&-2\xi_{12}	&2\xi^2_{13}&-2\xi_{13}	&-2\xi_{23}	&\xi^2_{12}		\cr
0			&-\frac[\xi_{02}/2]&\xi_{01}	&-\frac[\xi_{03}/2]&0			&\xi_{02}			\cr
\xi_{02}		&-\frac[\xi_{01}/2]&0		&0			&-\frac[\xi_{03}/2]&\xi_{02}			\cr
\xi_{03}		&0			&\xi_{03}	&-\frac[\xi_{01}/2]&-\frac[\xi_{02}/2]&0			\cr
\xi^2_{023}	&-\xi_{12}		&\xi_1^2	&-\xi_{13}		&0			&\xi_1^2			\cr
0			&\frac[\xi^2_{03}/2]&0	&-\frac[\xi_{23}/2]&-\frac[\xi_{13}/2]&\xi_{12}		\cr
\xi_2^2		&-\xi_{12}		&\xi^2_{013}&0			&-\xi_{23}		&\xi_2^2			\cr
0			&-\xi_{23}		&\xi_{13}	&\frac[\xi^2_{02}/2]&-\frac[\xi_{12}/2]&0			\cr
\xi_{23}		&-\frac[\xi_{13}/2]&0		&-\frac[\xi_{12}/2]&\frac[\xi^2_{01}/2]&0			\cr
\xi_3^2		&0			&\xi_3^2	&-\xi_{13}		&-\xi_{23}		&\xi^2_{012}		\cr
}\)
}{psymbol}
In fact, the rank of (\ShowLabel{psymbol}) is (of course) still 6, featuring the fact that the last 6 equations (which are rank 6 as well)
are sufficient to write a well-posed Cauchy problem for $\ga$.
The first 4 equations still do not contain second-order time derivatives: they are constraints on initial conditions.

Thus we split the original covariant equations into 4 constraint equations and 6 evolution equations for the bulk field $\ga^{ij}$.
We just have to show that the evolution problem is well-posed.
To do that we {\it can} show that the evolution system is symmetric hyperbolic, so that the Cauchy problem is well-posed.

The matrix of second order time derivatives (just set $\xi_0=1$ and $\xi_i=0$), given by
\eq{
\si(\xi) = -\Frac[1/4]\(\Matrix{
2	&0	&0	&0	&0	&0	\cr
0	&1	&0	&0	&0	&0	\cr
0	&0	&2	&0	&0	&0	\cr
0	&0	&0	&1	&0	&0	\cr
0	&0	&0	&0	&1	&0	\cr
0	&0	&0	&0	&0	&2	\cr
}\)
}
is in fact symmetric, non-degenerate and definite.
Unfortunately, the second space-derivatives and time-space-derivatives are not symmetric.
Too bad, but beautiful. Here is where one needs to resort to covariance (on $S$ this time, which is expressed by the momenta constraints).

We can split the non-symmetric evolution system 
\eqLabel{
e_{(ij)(jk)} \del_{00}\ga^{jk} + e^l_{(ij)(jk)} \del_{0l}\ga^{jk} + e^{lm}_{(ij)(jk)} \del_{lm}\ga^{jk} = b_{ij}
}{CovariantEvolution}
into its symmetric and shew-symmetric parts, namely
\eqLabel{
\Cases{
e_{(ij)(jk)} \del_{00}\ga^{jk} + e^l_{((ij)(jk))} \del_{0l}\ga^{jk} + e^{lm}_{((ij)(jk))} \del_{lm}\ga^{jk} = b_{ij}\cr
 e^l_{[(ij)(jk)]} \del_{0l}\ga^{jk} + e^{lm}_{[(ij)(jk)]} \del_{lm}\ga^{jk} =0\cr
}
}{NonCovariant evolution}
Of course, a solution of (\ShowLabel{NonCovariant evolution}) is a solution of (\ShowLabel{CovariantEvolution}), apparently not the other way around (which is what we need).
The splitting is not generally covariant! And the skew-symmetric part is a constraint. 
Believe it or not, non-covariance is what save us here.

\Note{
The skew-symemtric equation, just in view of its non-covariance, can be regarded as a constraint on coordinates on $S$.
It teaches us that there are coordinate systems on $S$ in which it is identically satisfied: they are called {\it harmonic coordinates}.
One can show that harmonic coordinates exist, and since equations (\ShowLabel{CovariantEvolution}) are covariant, it means that
whatever solution of it we consider, when written in harmonic coordinates, is a solution of the symmetric evolution equation (just since the skew part is identically satisfied).

In other words, it is just in view of non-covariance that we know that solution of (\ShowLabel{CovariantEvolution}) are exactly the solutions of the symmetric systems in harmonic coordinates, 
only written in an arbitrary coordinate system.
That is quite similar to the relation between geodesic motions and geodesic trajectories in relation to proper time and general parameterizations (see \cite{Geodesics}).
}

As a matter of fact, we find solutions $\ga^{ij}$ in harmonic coordinates, we interpret them as metrics $\ga$ on $S$ (in any coordinates) and, if we restrict to initial conditions which satisfy constraint equations
in the first place, we can rebuild a covariant metric $g$ on spacetime which is a solution of (vacuum) Einstein equations.

By considering the constraints as equations on space for the lapse $N$ and shift $\be$, the symbol in normal coordinates reads as
\eq{
\si(\xi) \propto \(\xi_1^2+ \xi_2^2+ \xi_3^2\)\(\Matrix{
1 & 0\cr
0&\ga_{ij}
}\)
}
which is manifestly non-degenerate. 
Hence, the constraints define an elliptic system {\it on space}.

We hope you appreciate as general covariance, constraint equations, evolution spatial equations interplay to account for under-determination as well as over-determination of field equations.
Of course, this analysis is equivalent to Hamiltonian canonical analysis, just based on covariant Lagrangian formalism. We do not mean it is better: it is more elementary though, more fundamental, namely near the fundamental issue of describing solution space.

From a quantum gravity perspective, it is natural to drop evolution equations:
that is precisely what LQG does. Maybe it is not the only way to do it, but for sure the result is a formalism which pastes together what one expects to be a quantum theory on a background free and covariant context.

\section*{Acknowledgements}

We acknowledge the contribution of INFN (Iniziativa Specifica QGSKY and Iniziativa Specifica Euclid), the local research project {\it  Metodi Geometrici in Fisica Matematica e Applicazioni (2023)} of Dipartimento di Matematica of University of Torino (Italy). This paper is also supported by INdAM-GNFM.
We are also grateful to S.Speziale and C.Rovelli for comments.

L. Fatibene would like to acknowledge the hospitality and financial support of the Department of Applied Mathematics, University of Waterloo, where part of this research was done.


\medskip


\begin{thebibliography}{99}

\bibitem{LN1}{L.Fatibene, A.Orizzonte, A.Albano, S.Coriasco, M.Ferraris, S.Garruto, N.Morandi,
{\it Introduction to Loop Quantum Gravity. The Holst's action and the covariant formalism},
Int. J. Geom. Meth. Mod. Phys. (accepted) 
{\tt https://doi.org/10.1142/S0219887824400164}
}

\bibitem{Rovelli1}{C.Rovelli, 
{\it Quantum Gravity}, 
Cambridge University Press, (2004) 
}

\bibitem{Rovelli2}{C.Rovelli, F.Vidotto, 
{\it An elementary introduction to Quantum Gravity and Spinfoam Theory},
Cambridge University Press, (2014)
}


\bibitem{Kolar}{I.Kol\'a\u r, J.Slov\'ak, P.W.Michor,
{\it Natural Operations in Differential Geometry},
Springer-Verlag Berlin Heidelberg (1993)
}

\bibitem{book1}{L.Fatibene, M.Francaviglia,
{\it Natural and Gauge-Natural theories},
Kluwer, Dordrecht, (2003)
}

\bibitem{Gelfand}{I.M.Gelfand, S.V.Fomin, 
{\it Calculus of Variations},
Prentice-Hall (1963)}

\bibitem{book2}{L.Fatibene,
{\it Relativistic theories, gravitational theories and General Relativity},
(unpublished) draft version 1.0.2.\\
{\tt http://www.fatibene.org/book.html}
}

\bibitem{Norton}{J.D.Norton, 
{\it General covariance and the foundations of general relativity: eight decades of dispute}, 
Rep. Prog. Phys. 56 (1993) 
}

\bibitem{HoleArgument}{L. Fatibene, M. Ferraris, G. Magnano,
{\it Constraining the Physical State by Symmetries},
Annals of Physics 378, (2017); 
{\tt arXiv:1605.03888}
}

\bibitem{GIMPSY}{M.J.Gotay, J.Isemberg, J.E.Marsden, R. Montgomery, J. \'Sniatycki, P.B. Yasskin,
{\it Momentum Map and Classical Relativistic Fields};
{\tt physics/9801019}
}

\bibitem{LN3}{L.Fatibene, A.Orizzonte,
{\it Lecture Notes in Loop Quantum Gravity. LN3: Boundary equations for Ashtekar-Barbero-Immirzi model}
(to appear)
}

\bibitem{Taylor}{M.Taylor, 
{\it Partial Differential Equations}, 
Springer, New York (1996)
}

\bibitem{Steenrod}{N.Steenrod,
{\it The Topology of Fibre Bundles}, 
PMS {\bf 14},
Princeton University Press, (1951);\goodbreak
{\tt https://www.jstor.org/stable/j.ctt1bpm9t5}
} 

\bibitem{Gar1}{L.Fatibene, S.Garruto,
{\it The Cauchy problem in General Relativity: An algebraic characterization},
CQG {\bf 32}(23) (2015); 
{\tt arXiv:1507.00476}
}

\bibitem{Choquet}{Y.Choquet-Bruhat, R.Geroch,
{\it Global aspects of the Cauchy problem in general relativity}, 
Comm. Math. Phys. {\bf 14},  (1969) 
} 

\bibitem{Gourgoulhon}{\'E.Gourgoulhon,
{\it 3+1 Formalism and Numerical Relativity},
General Relativity Trimester, Institut Henri Poincar\'e
}

\bibitem{Gar2}{L.Fatibene, S.Garruto,
{\it Principal Symbol of Euler-Lagrange Operators},
CQG {\bf 33}(14), (2016);\goodbreak
{\tt arXiv:1603.04732}
}

\bibitem{Geodesics}{L.Fatibene, M.Francaviglia,  G.Magnano,
{\it On a characterization of geodesic trajectories and gravitational motions},
Int. J. Geom. Meth. Mod. Phys. {\bf 09}(5),  (2012)	
}



\end{thebibliography}
 \end{document}